\documentclass[aps,prb,preprint]{revtex4}
\usepackage{amssymb,latexsym,amsmath}
\usepackage{graphicx}
\usepackage{color}
\usepackage{subfig}

\begin{document}

\title{Quantum Path Interference through Incoherent Motions in Multilevel Quantum Systems}
\author{Xin Chen}
\affiliation{Department of Chemistry,  \\ Massachusetts Institute of Technology, \\ Cambridge, MA 02139}

\date{now}
\begin{abstract}
Quantum path interferences or resonances in multilevel dissipative quantum systems play an important and intriguing role in the transport processes of nanoscale systems. Many previous minimalistic models used to describe the quantum path interference driven by incoherent fields are based on the approximations including the second order perturbation for the weak coupling limit, the ad-hoc choices of two-time correlation functions and $\it{etc}$. On the other hand, the similar model to study the non-adiabatic molecular electronic excitation have been extensively developed and many efficient quantum molecular dynamics simulation schemes, such as the Ehrenfest scheme, have been proposed. 

In this paper, I aim to construct an unified model, extend the Ehrenfest scheme to study the interactions of system-light and system-phonon simultaneously and gain insight into and principles of the roles of quantum path interferences in the realistic molecular systems. I discuss how to derive the time-dependent stochastic Schr$\ddot{o}$dinger equation from the Ehrenfest scheme as a foundation to discuss the detailed balance for the weak coupling limit and therefore the quantum correction in the Ehrenfest scheme.
Different from the master equation technique, the Ehrenfest scheme doesn't need any specific assumptions about spectral densities and two time correlation functions. 
With  simple open two-level and three-level quantum systems, I show the effect of the quantum path interference on the steady state populations. 
Currently I only focus on the role of the phonon thermal reservoir. The electromagnetic field (solar light) will be modeled as a thermal reservoir and discussed in detail in the future paper.

\end{abstract}

\maketitle
\newpage

\section{Introduction}

Having rigorous theories to model and study the quantum path interferences in open quantum systems is still a challenging task theoretically and computationally. The methods to determine and evaluate the interaction between a system and bath(reservoir) numerically and experimentally and simulate the evolution of a quantum subsystem are still not fully established. Using computational dynamic models which can take parameters from different electronic structure calculations and experimental results, I aim to have detailed understanding of quantum coherence/path interferences in realistic molecular systems.
Furthermore, I want to use the computational models to assist the synthesis and optimization of molecular systems/devices with strong quantum effects.

Recent theoretical studies based on minimalist master equation models\cite{scully,scully1,Das} show that the quantum path interference can induce absorption and emission cancellation and could play an interesting role in controlling the interfacial electron transfer in photovoltics\cite{sabas},  exciton transfer in the multi-chromophore molecular systems\cite{Dorfman} and singlet Fission \cite{zhu} which can increase the power conversion
efficiency beyond the so-called Shockley-Queisser limit .   
At the same time, the phonon can participate the non-radiative transitions (relaxation process) in molecular electronic systems and therefore the role of phonon is very similar to the one of the reservoir of vacuum oscillators in the Agarwal-Fano resonance. 
However the missing link between the minimalistic theoretical models and detailed atomistic understanding of these processes impedes the validation of these theoretical ideas and discoveries. I want to build computational models to address the dynamic influence from phonon and light\cite{pach} and study the quantum path interference in different transport and photo-chemical processes. Furthermore, the computational model can allow us to study the quantum resonance in solar cell including the realistic solar density of states and phonon spectral density. 

This paper consists of five sections: 1. in Section~\ref{model}, I review the Hamiltonian used to model the system-light and system-phonon interactions and propose an unified model ; 2. in Section~\ref{methodology}, I review the Ehrenfest scheme and the effective time-dependent stochastic Schr$\ddot{o}$rdinger equation for the unified model and Ehrenfest scheme;
3. in Section~\ref{detailed}, I review the concept of detailed balance and explain it within the Bloch-Redfield equation; 4. in Section~\ref{results}, I discuss the modified Ehrenfest wave-package propagation scheme including the detailed balance correction. I discuss the numerical results for three different cases to show how the quantum path interference lead to the detailed balance breakdown and manipulate the steady state population; 5. in Section~\ref{conclusion}, I present concluding remarks. 

\section{Unified Model}\label{model}
With a unified framework to describe the multi-level open quantum systems, I can describe the system-light and system-phonon interactions systematically and simultaneously, $e.g.$ the non-adiabatic radiative and non-radiative decay processes\cite{Kubo1}. 
The Hamiltonian for the system (matter) in interactions with the
reservoir of vacuum oscillators and incoherent field $R$  is defined as \cite{scully1}, 
\begin{equation}\label{scully}
H=H_r + H_R,
\end{equation}
where
\begin{equation}\label{hr}
H_r = H_0 + H_{rvo}+ H_{int} ,  
\end{equation}
where $H_0=\sum_{i=1}^N \epsilon_i \vert i \rangle \langle i \vert$, the interaction in the rotating wave approximation 
\begin{equation}
H_{int}=\sum_{i=1}^N\sum_{j=1}^{i-1} \sum_k g_k^{ij} \bigg (\hat{a}_k e^{-i \omega_k t} \vert i \rangle \langle j \vert + \hat{a}_k^{\dagger}   e^{i \omega_k t} \vert j \rangle \langle i \vert \bigg ),
\end{equation}
$g_k^{ij}= \frac{(\epsilon_i-\epsilon_j)\mu_{ij}}{\sqrt{2\varepsilon_0\omega_k V}}$ for the vacuum field modes\cite{agarwal},
and $H_{rvo}=\sum_k \hbar v_k \hat{a}^{\dagger}_k \hat{a}_k$ which can be ignored since the material system does not affect the light,
and 
\begin{equation}
H_R= \sum_{i=1}^{N}\sum_{j=1}^{i} \mu_{ij} \xi(t) \vert i \rangle \langle j \vert + H.c.,
\end{equation}
where $\xi(t)$ is a random process, $e.g.$ white noise. This Hamiltonian has been used to study population
trapping, lasing without inversion, and quenching of spontaneous emission via decays and incoherent pumping,
Aswagal-Fano resonance and interfacial electron ejection in quantum dots\cite{scully}. This model can be extended to define $g_k^{ij}$ for the interaction between the matter and solar light (instead of vacuum field). 

Similar to $H_r$,
the system-phonon Hamiltonian for the electronic excitation coupled to local phonon(Holstein Model)\cite{roden} is defined as, 
\begin{equation}\label{el}
H=H_{el}+H_{int}+H_{ph},
\end{equation}
where 
\begin{equation}
H_{el}=\sum_{ij}^N (\epsilon_i \delta_{ij}+J_{ij})\vert i \rangle \langle j \vert,
\end{equation}
\begin{equation}
H_{int}=  \sum_{i=1}^N  \sum_{j=1}^N\sum_k g_k^{ij}  \bigg (\hat{b}_{ijk} + \hat{b}_{ijk}^{\dagger} \bigg )  \vert i \rangle \langle j \vert ,
\end{equation}
and
\begin{equation}
H_{ph}= \sum_{j=1}^N \sum_k \hbar \omega_{jk} \hat{b}_{jk}^{\dagger}\hat{b}_{jk}.
\end{equation}
For the nonlocal phonon,
\begin{equation}
H_{int}=  \sum_{i=1}^N  \sum_{j=1}^N\sum_k g_k^{ij}  \bigg (\hat{b}_{k} + \hat{b}_{k}^{\dagger} \bigg )  \vert i \rangle \langle j \vert ,
\end{equation}
and
\begin{equation}
H_{ph}= \sum_k \hbar \omega_{k} \hat{b}_{k}^{\dagger}\hat{b}_{k}.
\end{equation}. In this model, I consider both energy fluctuation and the fluctuation in the energy transfer matrix elements $J_{ij}$. The fluctuation in the energy transfer matrix elements
is responsible for the quantum path interference through incoherent channels.

In general, the model of the multi-level Hamiltonian bilinearly coupled to Harmonic modes bath can be unified in terms of the following general matrix representation,
\begin{equation}\label{Ham}
    H= \sum_i \epsilon_i \vert i \rangle \langle i \vert + \sum_{i\neq j} V_{ij} \vert i \rangle \langle j \vert + \mathcal{C} + H_{rev}.
\end{equation}
For the Hamiltonian $H_r$ in Eq.~\ref{hr},
$\mathcal{C}_{ij} = (1-\delta_{ij} ) \sum_k g_k^{ij} \hat{a}_k \vert i \rangle \langle j \vert $ and $\mathcal{C}_{ji}=\mathcal{C}_{ij}^*$ and 
$H_{rev}= C_{rvo}$ which can be ingnored; and for the Hamiltonian in Eq.~\ref{el}, $\mathcal{C}_{ij}=\sum_k g_k^{ij} (\hat{b}_k + \hat{b}_k^{\dagger})$ for the nonlocal phonon and $H_{rev}=H_{ph}$.

Since the system and light interaction is weak, I can treat the evolution of quantum subsystem under the light with the second order perturbation. The Fermi golden rule or master equation can be used to calculate the influence of light on the evolution of quantum subsystems. However, for the phonon reservoir, I need a better treatment because phonon can't be ignored and interaction is ofter is not perturbation. As well, the phonon reservoir with specific spectral density can replace the incoherent field $H_R$ in Eq.~\ref{scully}. The Hamiltonian used to study the dynamics of nanoscale systems interacting with the thermal reservoirs of light and phonon can be defined as,
\begin{equation}\label{NanoHam}
    H= \sum_i \epsilon_i \vert i \rangle \langle i \vert + \sum_{i\neq j} V_{ij} \vert i \rangle \langle j \vert + \mathcal{C}_{light} + \mathcal{C}_{ph} + H_{ph}.
\end{equation}
The coefficient can be obtained computationally and numerically\cite{bredas}. In the rest of the paper, I will focus on $\mathcal{C}_{ph}$. The treatment of $\mathcal{C}_{light}$ can be studied independent of $\mathcal{C}_{ph}$ using the effective Hamiltonian \cite{kato1,kato2} based on the Fermi golden rule or perturbative Master equation derived in reference \cite{scully1} or even computationally. It will be discussed in the future.

\section{Ehrenfest Mixed Quantum-Classic Dynamics and Stochastic Schr$\ddot{O}$dinger Equation}\label{methodology}

Give this high dimensional complex Hamiltonian in Eq.~\ref{Ham}, it is normally impossible to simulate the whole dynamic evolution with full quantum mechanic description. Different versions of mixed quantum-classic schemes \cite{john,kapral} in Schr$\ddot{o}$dinger or Liouville space are often used by treating phonon with classical mechanics and quantum subsystem with quantum mechanics to reduce the complexity of numerical simulations.   
Among the schemes, 
the Ehrenfest scheme is a  popular choice to simulate the evolution of a quantum subsystem described with Eq.~\ref{Ham}. 

For the following discussion, I assume $g_k^{ij}$  are the same for all the $(i,j)$ pairs. Therefore, $\mathcal{C}_{ph}=V \times Q$ where $Q=\sum_k g_k (\hat{b}_{k} + \hat{b}_{k}^{\dagger}) $.
The total wave-function in the Ehrenfest scheme is assume to be factorized into a product of the  subsystem and individual modes,
\begin{equation}\label{eh1}
\psi(S,q_1,q_2,\cdots,q_N,t)\approx\phi(S,t) \times \prod_{i=1}^N \xi_i (q_i,t),
\end{equation}
where $S$ is the energy eigenbasis of the quantum subsystem, $q_i$ and $p_i$ are the dimensionless position and momentum of a Harmonic mode in phonon \cite{sakurai} ($\hat{q}_i=\sqrt{\frac{\hbar}{2}} (\hat{b}_{i}  + \hat{b}_{i}^{\dagger})$ position operator of a Harmonic mode,
and its conjugate momentum operator, $\hat{p}_i=i\sqrt{\frac{\hbar}{2}}(\hat{b}_{i}^{\dagger}-\hat{b}_{i})$).
The evolution of the wave-function of the quantum subsystem  $\phi(S,t)$ can be expressed as,
\begin{equation} \label{sys_eom}
i\hbar \; \frac{\partial \phi(S,t)}{\partial t} = H_S \; \phi(S,t),
\end{equation}
where
\begin{equation}\label{sysh}
H_S =  H_0+{V}\times Q(t),
\end{equation}
$H_0 = \sum_i \epsilon_i \vert i \rangle \langle i \vert + \sum_{i\neq j} J_{ij} \vert i \rangle \langle j \vert$, $V=\sum_{ij} V_{ij}  \vert i \rangle \langle j \vert$ where $V_{ij}$ is a coupling constant and $Q(t)=\sum_i g_i q_i(t)$. For the clarification, I consider the single phonon case in the discussion, $\it{i.e.}$ $X= \sum_i \sum_j \sum_k g_k (\hat{b}_{k}  + \hat{b}_{k}^{\dagger}) \vert i \rangle \langle j \vert $ in Eq.~\ref{Ham}.  I want to emphasize that $H_S$ is different from, but maybe equivalent to in some way, the stochastic Hamiltonian used in the Gauss-Markov model\cite{jackson}. 
Correspondingly,
the equations of motion for the individual mode, $(q_i(t),\;p_i(t))$ can be expressed as,
\begin{eqnarray} \label{eom_bath}
\frac{dq_i}{dt} & = &  \frac{\partial \mathcal{H}_{phe}(t)}{\partial p_i}, \nonumber \\
\frac{dp_i}{dt} & = & -\frac{\partial \mathcal{H}_{phe}(t)}{\partial q_i} ,
\end{eqnarray}
where
\begin{equation} \label{phononh}
\mathcal{H}_{phe}(t)=\sum_i\frac{p_i^2}{2} + \frac{1}{2} \omega_i^2 q_i^2  +  g_i z_i(t) q_i   ,
\end{equation}
where $z_i(t)=\frac {\partial}{\partial q_i} \langle   \phi(S,t) \vert {V} \times Q \vert  \phi(S,t) \rangle  $  is time dependent determined by $\vert  \phi(S,t) \rangle$ since the coupling is bilinear ${V}\times Q$. $z_i q_i $ is the time dependent influence from the quantum subsystem on the individual modes.

In order to kick out the simulation, the thermal Wigner function is used in the Ehrenfest scheme to generate the initial configurations,
\begin{equation}\label{wigner}
W(q_i(0),p_i(0))=2^N\prod_{i=1}^N  \text{tanh}(h_i/2) \exp \left (-\text{tanh}(h_i/2)(\frac{ \omega_i}{\hbar} q_i(0)^2+ \frac{1}{\omega_i \hbar} p_i(0)^2) \right),
\end{equation}
where $h_i = \hbar  \beta \omega_i$. I can sample the distribution function in the phase space $(Q(0), P(0)$ where $Q(0) = {q_1(0),q_2(0),\cdots,q_i(0)\cdots q_N(0)}$, $P(0) = {p_1(0),p_2(0),\cdots,p_i(0)\cdots p_N(0)}$, and calculate the evolution of the dynamic trajectories $(Q(t),P(t))$ for all configurations according to Eq.~\ref{eom_bath}.
The observables of the quantum subsystem can be evaluated as $\langle \hat{O} (t) \rangle =\frac{1}{M} \sum_{j=1}^M \langle \phi(S,t|Q_j(t),P_j(t)) \vert \hat{Q}  \vert \phi(S,t| Q_j(t),P_j(t)) \rangle $, where $Q_j(t)$ and $P_j(t)$ are the $j\it{th}$ configuration and $M$ is the total number of configurations.

\subsection{Implied Time-dependent Stochastic Schr$\ddot{o}$dinger Equation}
In this subsection, I want to show that the evolution of the quantum subsystem based on the Ehrenfest scheme can be reduced to an equivalent time-dependent Stochastic Schr$\ddot{o}$dinger equation\cite{gaspard} (a quantum Langevin equation in Schr$\ddot{o}$dinger picture). In the Liouville space, the operator quantum Langevin equation (for example, quantum Master Equation for reduced density matrix) can be derived using the  Nakajama-Zwanzig projector technique for the weak coupling limit. I want to point out that the two methods are essentially equivalent in mathematics for the high temperature limit, $\it{i.e.}$, a thermal reservoir can be treated as a classical color noise with a classical time correlation function. How to derive the equivalence of the two methods will be not the topic of this paper. But it is briefly shown in the references \cite{gaspard} and others for the weak coupling limit.

For the classic
system-bath Hamiltonian, 
\begin{equation}
H= \frac{P_S^2}{2M} + V(X_S)  +  \sum_i g_i q_i  {Q} (X_S)+  \sum_i \frac{p_i^2}{2m_i} + \frac{1}{2} m_i \omega_i^2 q_i^2 ,
\end{equation}
where $\frac{P_S^2}{2M} + V(X_S)$ is the system Hamiltonian and $Q(X_S)$ is the function of system coordinate $X_S$, the equation of motion of the system can be expressed as,
\begin{eqnarray}\label{cle}
\frac{dX_S(t)}{dt} & = & P_S (t)/M; \\ \nonumber
\frac{dP_S(t)}{dt} & = & -\frac{dV(X_S)}{dX_S} - \frac{dQ(X_S)}{dX_S} \sum_i g_i q_i,   
\end{eqnarray}
and the evolution of an individual mode can be expressed as,
\begin{equation}
q_i(t)= q_i(0) \cos(\omega_i t) + \frac{1}{m_i \omega_i} p_i(0) \sin(\omega_i t) - \frac{g_i}{m_i \omega_i  }
 \int_0^t ds \sin(\omega_i (t-s)) Q(s).
\end{equation} 
and using the integration by parts, the equivalent form can be expressed as,
\begin{equation}
q_i(t)=q_i(0) \cos(\omega_i t) + \frac{1}{m_i\omega_i} p_i(0) \sin(\omega_i t)  - \frac{g_i}{m_i \omega_i^2}
 \left [ Q(t)-Q(0)\cos(\omega_i t)-\int_0^t ds \cos(\omega_i (t-s)) \frac{dQ(s)}{dt} \right ].
\end{equation}
The equation of motion of the system can be re-written as the classical generalized Langevin equation \cite{zwangzig},
\begin{eqnarray}\label{cle}
\frac{dX_S(t)}{dt} & = & P_S (t)/M; \\ \nonumber
\frac{dP_S(t)}{dt} & = & -\frac{dV(X_S)}{dX_S} -\frac{dQ(X_S)}{dX_S} \bigg (\int_0^t ds K(t-s) \frac{dQ(s)}{ds} -  K(0) Q(t)  + \mathcal{F}(t) \bigg),
\end{eqnarray}
where
$K(t)=\sum_i \frac{g_i^2}{m_i\omega_i^2} \cos(\omega_i t)$, memory friction kernel and
$\mathcal{F}(t)$ is the fluctuating force,
\begin{equation} \label{fluc}
\mathcal{F}(t) = \sum_i g_i \left (q_i(0) +\frac{g_i  {Q}(0)}{m_i \omega_i^2} \right) \cos(\omega_i t) + \frac{g_i p_i(0)}{m_i \omega_i^2} \sin(\omega_i t),
\end{equation}
where $\langle \mathcal{F} (t) \rangle =0$ and $\langle \mathcal{F}(t) \mathcal{F}(0) \rangle = k_{\beta} T K(t)$ using the thermal average in the
initial state of the reservoir with the shifted canonical equilibrium distribution\cite{weiss}. 

Similar to the classic generalized Langevin equation,  the time-dependent stochastic  Schr$\ddot{o}$dinger Equation\cite{ram} can be derived based on the Ehrenfest scheme in Eqs.~\ref{sys_eom} and \ref{eom_bath}, 
\begin{equation}\label{ess}
i\hbar \; \frac{\partial \phi(S,t)}{\partial t} = (H_0+{V}\times Q(t))\; \phi(S,t),
\end{equation}
where the environment fluctuation is defined as,
\begin{equation}
Q(t)=\sum_i g_i \bigg [ q_i(0) \cos(\omega_i t) + \frac{1}{\omega_i} p_i(0) \sin(\omega_i t) \bigg ] - \frac{g_i^2}{\omega_i^2}
 \left [ z_i(t)-z_i(0)\cos(\omega_i t)-\int_0^t ds \cos(\omega_i (t-s)) f_i(s) \right].
\end{equation}
where $z_i(t)$ is the time-dependent displacement and $f_i(t)=\frac{dz_i(t)}{dt}$ the effective velocity ($m_i=1$ given that $p_i$ and $q_i$ are dimensionless). 
Therefore
the time-dependent stochastic Schroedinger equations for the quantum subsystem can be expressed as,
\begin{equation} \label{qle}
i\hbar \; \frac{\partial \phi(S,t)}{\partial t} = \left [ H_0+ V \times \left (\int_0^t ds K(t-s) f_i(s)  -K(0) z_i(t) + \mathcal{F}(t) \right) \right ] \; \phi(S,t),
\end{equation}
where
\begin{equation}
\mathcal{F}(t)= \sum_i g_i \left (q_i(0) +  \frac{g_i z_i(0)}{\omega_i^2} \right) \cos(\omega_i t) + \frac{g_i p_i(0)}{\omega_i^2} \sin(\omega_i t),
\end{equation}
is equivalent to Eq.~\ref{fluc} and therefore
the kernel $\langle \mathcal{F}(t) \mathcal{F}(0) \rangle = k_{\beta} T K(t) $. When the memory kernel becomes a delta function, this model is reduced to the Caldeira-Leggett model \cite{cald} (quantum Brownian motion). 

\subsection{Noise and Spectral Density}
The time-dependent stochastic Schr$\ddot{o}$dinger shows that the quantum state $\phi(t)$ evolves under the classical Gaussian color noise  $\mathcal{F}(t)$.
The noise is characterized by the classical time correlation function $C_{cl}(t)=\langle \mathcal{F}(t)\mathcal{F}(0) \rangle $ is even and symmetric, $\it{i.e.}$, $C_{cl}(t)= C_{cl}(-t)$. The solutions to the time-dependent stochastic Schr$\ddot{o}$dinger for some specific cases, such as Ornstein-Ulenbeck, have been discussed\cite{Fox}.

The coupling coefficients in the bilinear coupling, $g_i$, determine the nature of noise and the dissipative dynamics. They can be evaluated computationally or empirically\cite{bredas,kato1}. On the other hand, in theoretical models, the spectral density involving the $g_i$ coefficients and frequencies,
\begin{equation}
J(\omega)= \frac{\pi}{2} \sum_i \frac{g_i^2}{\omega_i } \delta(\omega-\omega_i),
\end{equation}
are used to define the memory kernel $K(t)=\frac{2}{\pi} \int d\omega J(\omega) cos(\omega t)/\omega$, the time correlation function of noise,
and the reduced dynamics of the quantum subsystem\cite{xin}.  
Some popular forms of spectral densities is continuous function, such as ohmic with exponential cutoff $\eta \omega e^{-\omega/\omega_c}$ and Drude ohmic with Lorentzian cutoff $2\eta \omega_c \frac{\omega}{\omega^2+\omega_c^2}$. In order to simulate these kinds of spectral densities, discretization schemes\cite{berkelbach,moix,wang} are needed to obtain $g_i$ and $\omega_i$ . For example, the exponential ohmic spectral density can be discretized as,\cite{moix} 
\begin{equation}
\omega_i  =  - \omega_c \log \left [  1-\frac{i}{N} \left( 1-\exp(-\frac{\omega_m}{\omega_c}) \right ) \right ],
\end{equation}
and
\begin{equation}
g_i  =   \omega_i \sqrt{\frac{2\eta}{\pi}\frac{\omega_c}{N} \left[ 1-\exp(-\frac{\omega_m}{\omega_c}) \right]},
\end{equation}
which will be used in the calculation in the Section~\ref{results}.
The number of modes should reproduce the reorganization energy $\mu=\frac{1}{\pi} \int_0^\infty d\omega J(\omega)/\omega$, $\it{i.e.}$ $\frac{1}{2}\sum_i \frac{g_i^2}{\omega_i^2}\approx \frac{1}{\pi} \int_0^\infty d\omega J(\omega)/\omega $. 

However, I want to emphasize that the covariance decomposition method can be used to generate the Gaussian noise with arbitrary spectral densities including both discrete and continuous spectral densities\cite{xin}.

\section{Detailed Balance in Open Quantum Systems and Steady State Equilibrium} \label{detailed}

The detailed balance conditions in open quantum systems have been discussed and established in literatures\cite{qdb1,qdb2,qdb}. The concept of the detailed balance is associated with the  quantum two-time  correlation function and the weak-coupling limit of interaction or the Markovian limit of the correlation time\cite{xinbob}. 
For the classical system, the detailed balance \cite{klein,htheorem} has the following linear relationship of the kinetic rate
\begin{equation}
k_{i\leftarrow j} \exp(-\beta \epsilon_j) =k_{j\leftarrow i} \exp(-\beta \epsilon_i),
\end{equation}
in the master (linear kinetic) equation,
\begin{equation}
\frac{dp_i}{dt} = \sum_j (k_{j\leftarrow i} p_j-k_{i\leftarrow j} p_i).
\end{equation}

For the quantum master (kinetic) equation, the detailed balance is reflected in the Fourier transform of two time quantum correlation function, $C(\omega)=e^{\beta \hbar \omega}C(-\omega)$ 
(or in time domain, $C(t)=C^*(-t)$ and its periodic condition $C(t)=C^*(t-i\beta \hbar )$)
 \cite{mark,nitzan}. This relationship apparently doesn't hold in the Ehrenfest scheme which has classical two-time correlation function. 

In this paper, I use the Bloch-Redfield equation to discuss the concept of quantum detailed balance. The complete description of quantum detailed balance beyond the weak coupling limit is still not fully established and will be an important future theoretic task. The connection between the master equation and the time-dependent Schr$\ddot{o}$dinger equation is discussed in the reference\cite{gaspard} for the second order limit. 
The discussion of the quantum detailed balance correction for the classic time correlation function in the Ehrenfest scheme will be postpone to Section~\ref{results}.

\subsection{Detailed Balance and Bloch-Redfield Equation}

The evolution of the reduced density matrix can be expressed in terms of the infinite summation of multi-time correlation function (memory kernels) according to the cumulant expansion technique\cite{kubo} and Nakajima-Zwangzig projection operator technique\cite{Nakajima,zwangzig}.   
After truncating the summation of multi-time memory kernels at the second order,  two different time ordering prescriptions can be obtained: partial time ordering prescription (POP) and chronological time ordering prescription (COP)\cite{deutch,Mukamel,mukamel1}. As a result, two kinds of the second-order master equations (rate equation)\cite{localnon} can be obtained, the time-local convolutionless second order master equation for the POP case;
and the time-nonlocal convolution  second order master equation for the COP case. In general, the second-order master equation is governed by the quantum two-time correlation function \cite{Makowski,Makowski1} by sacrificing the complete description of the time-ordering multi-time correlation functions (memory kernel) due to the truncation \cite{Makowski,deutch,Mukamel,mukamel1}. 

The Bloch-Redfield equation can be derived from either the COP or POP master equation in the eigenbasis of quantum subsystem Hamiltonian.
For the Hamiltonian of $H_0+V\times Q +H_{ph}$, 
the Bloch-Redfield master equation\cite{redfield,nitzan,suarez} is expressed as,
\begin{eqnarray} \label{Redfield}
\frac{d\rho_{ij}}{dt}&=&-\frac{i}{\hbar} (\epsilon_{i} -\epsilon_{j})\rho_{ij} -\frac{i}{\hbar}(J_{ik}\rho_{kj}-\rho_{ik}J_{kj})\\ \nonumber
&& -\sum_{kl} \bigg ( R_{ik,kl}(\omega_{lk}) \rho_{lj} + R^*_{jl,lk}(\omega_{kl}) \rho_{ik} \\ \nonumber
&&  -[R_{lj,ik} (\omega_{li}) +R_{ki,jl}^* (\omega_{lj})] \rho_{kl} \bigg )
\end{eqnarray}
where
\begin{equation}
R_{ij,kl}(\omega)= \frac{1}{\hbar^2}  \int_0^{\infty} C(t) \exp(i\omega t) V_{ij} V_{kl},
\end{equation}
where $C(t)=\langle Q (t) Q (0)\rangle$ is an assumption to the Bloch-Redfield equation which has to be defined in an adhoc way. However for the time-dependent stochastic Schr$\ddot{o}$dinger equation, the $C(t)$ is intrinsically determined by $Q(t) = \int_0^t K(t-s) f_i(t)  -K(0) z_i(t) + \mathcal{F}(t)$

It is clear that the detailed balance have the binary connection solely associated with two energy levels, which is determined by the weak coupling and second order perturbation.  However, the standard Ehrenfest propagation scheme doesn't have detailed balance constraint and leads to the high temperature equal distribution steady state due to the classical time correlation function, $\it{i.e.}$ $\exp{-\beta \omega_{ij}}=1$ when $\beta \rightarrow 0$ at the high temperature. I assume that the imaginary parts of quantum correlation functions goes to zero at the high temperature. 

In the next subsection, I will use a two-level system as an example to elaborate how the detailed balance is enforced in the second order Bloch-Redfield master equation (weak coupling limit). I want to emphasize that the Block-Redfield equation is very similar to the one used by Harris and Scully \cite{harrisa} to study the Fano-like quantum path interference.

\subsubsection{Two-Level Model}
The Bloch-Redfield master equation essentially is a quantum version kinetic rate equation. I take a two level system as an example,
\begin{equation}
H=H_0+ V \times Q,
\end{equation}
\begin{equation}
H_0=H_S+H_{ph},
\end{equation}
\begin{equation}
H_S=
\left [
\begin{array}{cc}
\epsilon_1 & 0 \\
0 & \epsilon_2
\end{array}
\right]
,
\end{equation}
and
\begin{equation}
{V}=
\left[ \begin{array} {cc}
0 & 1 \\
1 & 0
\end{array} \right ].
\end{equation}
In this Hamiltonian, I only turn on the off-diagonal incoherent channels, $V_{12}$ and $V_{21}$, $\it{i.e.}$ energy relaxation channels; turn off the diagonal incoherence channels, $\it{i.e.}$ $V_{11}=0$ and $V_{22}$, $\it{i.e.}$ energy dephasing channels. Also, the coherent transition channels,  $J_{12}=J_{21}=0$, are turned off.

The corresponding Block-Redfield master equation\cite{nitzan}  is defined as,
\begin{eqnarray} \label{db1}
\frac{d\rho_{11}}{dt} & = & -2  \text{Re} R_{12,21}(\omega_{12}) \rho_{11} + 2 \text{Re} R_{21,12}(\omega_{21}) \rho_{22}, \\ \label{db2}
\frac{d\rho_{22}}{dt} & = & -2 \text{Re} R_{21,12}(\omega_{21}) \rho_{22} + 2 \text{Re} R_{12,21}(\omega_{12}) \rho_{11},\\ \nonumber
\frac{d\rho_{12}}{dt} & = & -i \omega_{12} \rho_{12} -[R_{12,21}(\omega_{12}) +R_{21,12}^*(\omega_{21})]\rho_{12} +[R_{12,12}(\omega_{21}) + R_{21,21}^*(\omega_{12})] \rho_{21} \\ \nonumber
&& +[R_{11,21}(\omega_{12}) -R_{22,21}^*(\omega_{12})]\rho_{11} +[R_{22,12}(\omega_{21}) - R_{11,12}^*(\omega_{21})] \rho_{22} ,
\end{eqnarray}
where $\text{Re} R_{12,21}(\omega_{12})=\frac{1}{\hbar^2} V_{12}V_{21}\int_{-\infty}^{\infty} dt e^{-i\omega_{12}t} C(t) $ and $\text{Re} R_{21,12}(\omega_{21}) = \frac{1}{\hbar^2} V_{21}V_{12} \int_{-\infty}^{\infty} dt e^{-i\omega_{21}t} C(t)$. For this model, $[R_{11,21}(\omega_{12}) -R_{22,21}^*(\omega_{12})]\rho_{11} +[R_{22,12}(\omega_{21}) - R_{11,12}^*(\omega_{21})] \rho_{22}$ will disappear since the energy dephasing channels, $V_{11}$ and $V_{21}$, are turned off.
The quantum detailed balance condition in the Bloch-Redfield equation, $C(\omega_{ij})=\exp(-\beta \hbar \omega_{ij}) 
C(\omega_{ji})$ due to the properties of the quantum time correlation function,  can be mapped to be $k_{2\leftarrow1}=2\text{Re}R_{12,21}(\omega_{12})$ and $k_{1\leftarrow 2}=2\text{Re}R_{21,12}(\omega_{21})$.

\section{Modified Ehrenfest Propagation Scheme with Detailed Balance Correction}
$Q(t)$ can be considered as the fluctuation induced by the harmonic thermal reservoir.  For the Ehrenfest scheme in Eq.~\ref{qle}, 
$Q(t)=\int_0^t K(t-s) f_i(t)  -K(0) z_i(t) + \mathcal{F}(t) $ and therein
the time correlation function is even, symmetric and real-valued  $C_{cl}(t)=C_{cl}(-t)$. I discussed in Section~\ref{detailed} that  in the Bloch-Redfield equation, $C(t)$ is the important input to the equations. However, the Ehrenfest scheme doesn't need  $C(t)$ as the input since Eqs.~\ref{eom_bath} gives the dynamic evolution of $Q(t)$ without the enforcement of detailed balance. In the following part, I will show how to make the detailed balance correction suggested by the Bloch-Redfield equation.

I want to emphasize that at the high temperature limit, the quantum time correlation function $C(t)$ will be reduced to the classical time correlation function $C_{cl}(t)$ since the imaginary part of $C(t)$ becomes zero\cite{kubo}. The quantum detailed balance is embed in the imaginary part of $C(t)$. Fixing the Ehrenfest scheme is in an ad-hoc way to consider the effect of imaginary part of $C(t)$.

For the time-dependent stochastic Schroedinger Equation derived from the Ehrenfest scheme, I have to modify the Hamiltonian to enforce the relationships in Eqs.~\ref{db1} and ~\ref{db2}. The connection between the thermal rate $k_{i\leftarrow j}$
and non-equilibrium Fermi gold rule's rate is revealed through the Fourier transform of $C_t$, $C(\omega_{12})$ and $C(\omega_{21})$
\begin{eqnarray}\label{redfield}
k_{2\leftarrow 1}=2\text{Re}R_{12,21}(\omega_{12}) & = &  \frac{2\pi}{\hbar^2} \vert \langle 2\vert   {V} \vert 1\rangle \vert ^2 C(\omega_{12}),  \\ \nonumber
k_{1\leftarrow 2}=2\text{Re}R_{21,12}(\omega_{21}) & = &  \frac{2\pi}{\hbar^2} \vert \langle 1\vert   {V} \vert 2\rangle \vert ^2 C(\omega_{21}),
\end{eqnarray}
where is  $\langle 2\vert \mathcal{V} \vert 1\rangle $ and $ \langle 1\vert \mathcal{V} \vert 2\rangle $  are the off-diagonal matrix elements in the coupling matrix $V$ in the energy eigenbasis.

For the harmonic bath, $C(\omega)=\frac{1}{1+ 2\exp(- \beta \hbar \omega)} C_{cl}(\omega)$. Therefore, I have the approximations, $C(\omega_{12})=\frac{1}{1+ 2\exp(- \beta \hbar \omega_{12})}  C_{cl}(\omega_{12})$ and
$C(\omega_{21})=\frac{1}{1+ 2\exp(- \beta \hbar \omega_{21})}  C_{cl}(\omega_{21})$. As a result, I can include the quantum correction factor in the effective Hamiltonian for the time-dependent Schr$\ddot{o}$dinger equation\cite{skinner} by modifying ${V}$ to $V^m$ in which $V_{12}^m=\langle 1 \vert {V}^m \vert 2\rangle  = \left ( \frac{1}{1+ 2\exp(- \beta \hbar \omega_{12})}  \right)^{1/2}$ and $V_{21}^m= \langle 2\vert   {V}^m \vert 1\rangle  = \left ( \frac{1}{1+ 2\exp(- \beta \hbar \omega_{21})}  \right)^{1/2}$. The transition probability in the Schr$\ddot{o}$dinger picture $k_{2\leftarrow 1} \propto {V_{12}^m}^2 C_{cl}(t) $ and $k_{1\leftarrow 2} \propto {V_{21}^m}^2 C_{cl}(t)$ where $C_{cl}(t)= \langle Q(t) Q(0) \rangle_{ph}$ according to the second order perturbation Fermi golden rule and the average over the configurations of initial states. 
This detailed balance correction scheme has been suggested by some previous work\cite{Bastida,peskin,aghtar} in different context.

Therefore, the new equation of motion of the quantum subsystem according to the modified Ehrenfest scheme is:
\begin{equation} \label{sys_eom_mod}
i\hbar \; \frac{\partial \phi(S,t)}{\partial t} = H_S^m \; \phi(S,t),
\end{equation}
where
\begin{equation}
H_S^m =  H_0+\mathcal{V}^m\times Q(t),
\end{equation}
where the matrix elements in $\mathcal{V}^m$ is $V_{ij}^m$ as defined previously.
For the reservoir, nothing is changed, $\it{i.e.}$ the original coupling matrix $\mathcal{V}$ and $z_i(t)$ are used. In addition, I want to mention that I only correct the detailed balance of population part. Instead, the relationship reflected in the coherence part, $[R_{11,21}(\omega_{12}) -R_{22,21}^*(\omega_{12})]\rho_{11} +[R_{22,12}(\omega_{21}) - R_{11,12}^*(\omega_{21})] \rho_{22}$, is neglected.

The propagation scheme for the individual configuration has four steps:
\begin{enumerate}
\item Evaluate the effective time-dependent Hamiltonian for the quantum subsystem,
\begin{equation}
H_S^e(t)=H_0+V^m \times Q(t),
\end{equation}
where $Q(t)=\sum_k g_k q_k(t)$.
\item Propagate the quantum subsystem,
\begin{equation}
\phi(S,t+dt) = \exp (-i H_S^e(t) dt ) \phi (S,t).
\end{equation}
\item 
Evaluate the effective Hamiltonian for the phonon,
\begin{equation}
\mathcal{H}_{ph}^e(t)=\sum_i\frac{p_i^2}{2} + \frac{1}{2} \omega_i^2 q_i^2  +  \langle \phi(t)\vert V \times Q(t) \vert \phi(t) \rangle;
\end{equation}
and for the individual model, the effective Hamiltonian is,
\begin{equation}
H_i = \frac{p_i^2}{2} + \frac{1}{2} \omega_i^2 q_i^2  +  g_i z_i(t) q_i.
\end{equation}
\item Propagate the individual mode in phonon with the Verlet algorithm\cite{verlet} according to Eq.~\ref{eom_bath},
\begin{eqnarray}
q_i(t+dt)=  q_i(t) + p_i(t) dt + \frac{1}{2} f(t) dt^2, \\ \nonumber
p_i(t+dt)=  p_i(t) + \frac{f(t)+f(t+dt)}{2}  dt, 
\end{eqnarray}
where $f(t)=\omega_i^2 q_i(t) + g_i z_i(t) q_i(t) $.
\end{enumerate}
In order to kick out the propagation scheme, I need to sample the configurations of initial states $(q_i(0) , p_i(0))$ according to Eq.~\ref{wigner}. The observables of the quantum subsystem can be evaluated as $\langle \hat{O} (t) \rangle =\frac{1}{M} \sum_{j=1}^M \langle \phi(S,t|Q_j(t),P_j(t)) \vert \hat{Q}  \vert \phi(S,t| Q_j(t),P_j(t)) \rangle $ where $M$ is the number of configurations. The matrix element of the reduced density matrix  can be evaluated using the projection operator $\hat{P}_{ij}=\vert i \rangle \langle j \vert $.

\section{Simulation Results and Discussion}\label{results}

\begin{figure}
\includegraphics[width=6in]{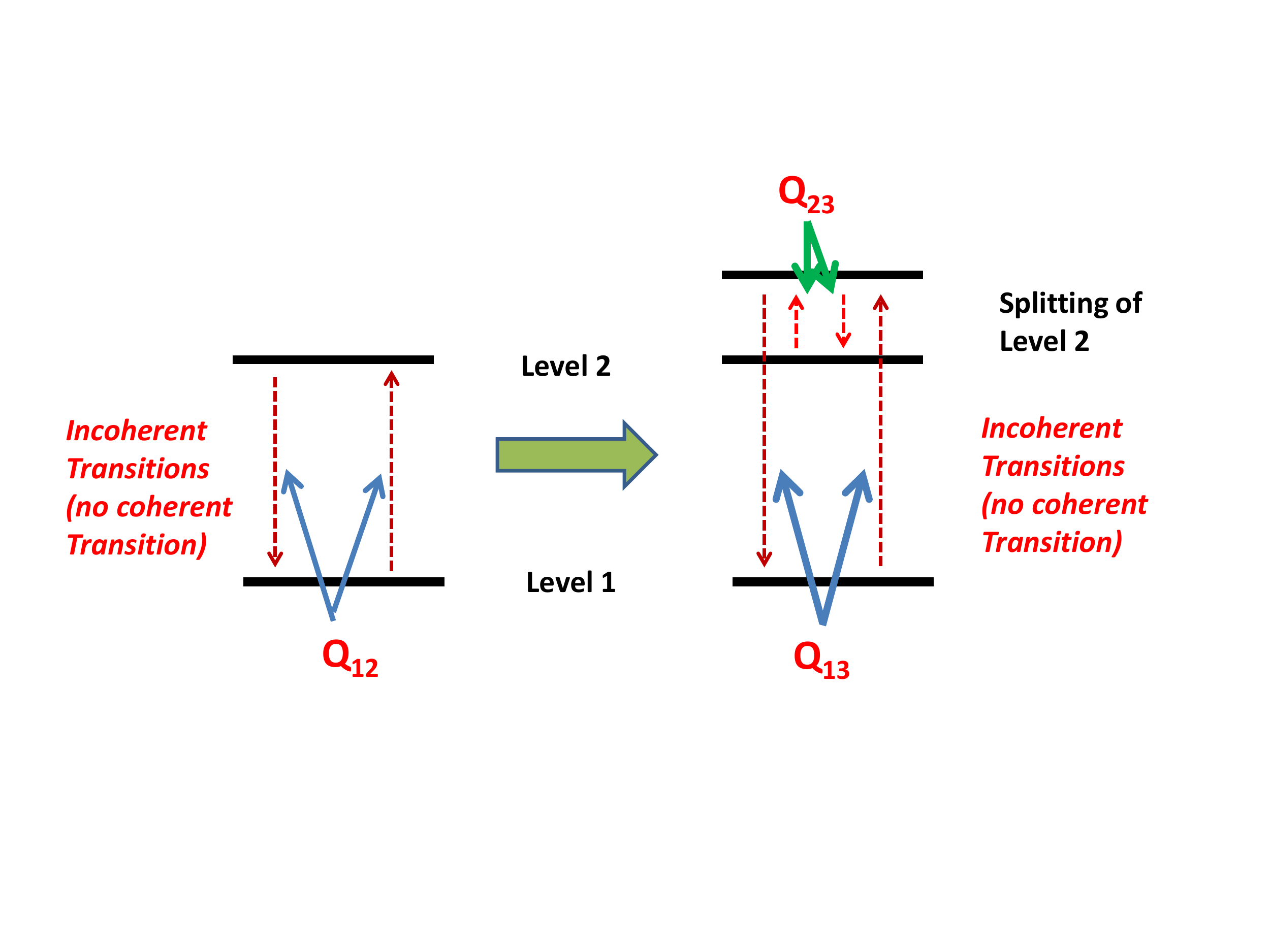}
\caption{In this diagram, I show that the setup of the two level model and three level models used in the numerical calculations. $V_{ij}$ is the matrix element in the matrix $V$ in Eq.~\ref{ess} }\label{diagram}
\end{figure}

In this section, I elaborate the modified Ehrenfest method based on the setups of two level and three level quantum systems shown in the diagram presented in Figure~\ref{diagram}.
In the following subsection, I will discuss three different scenarios: A two level system coupled to a single thermal reservoir in subsubsection~\ref{case1}; 2. A three level system coupled to one reservoir in subsubsection~\ref{case2}; 3. A three Level system coupled to two thermal reservoirs in subsubsection~\ref{case3}.

\subsubsection{Two Level System Coupled to a Single Thermal Reservoir}\label{case1}
\begin{figure} \begin{tabular}{cc}
\subfloat[]{\includegraphics[width=3in]{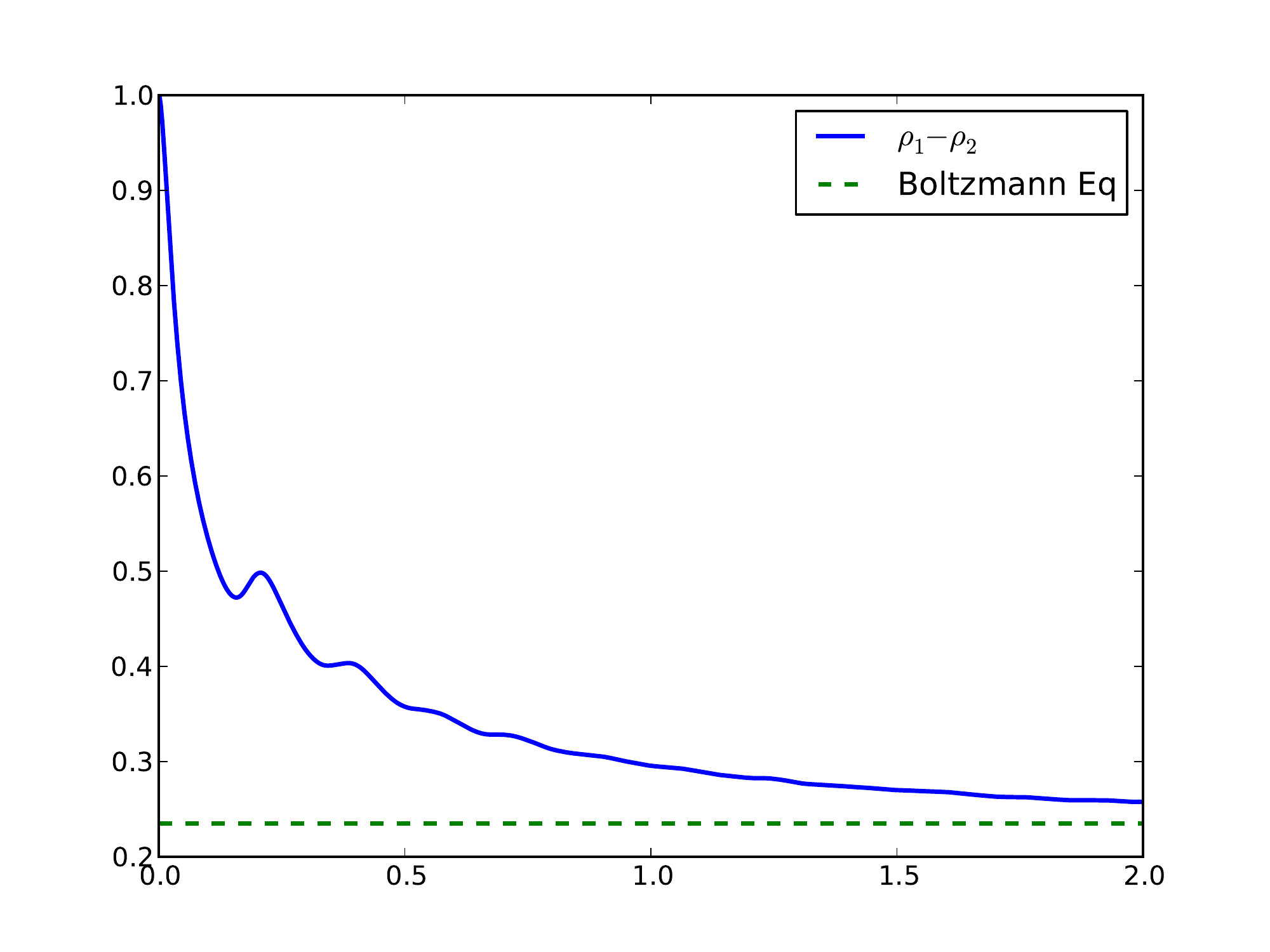}}
   & \subfloat[]{\includegraphics[width=3in]{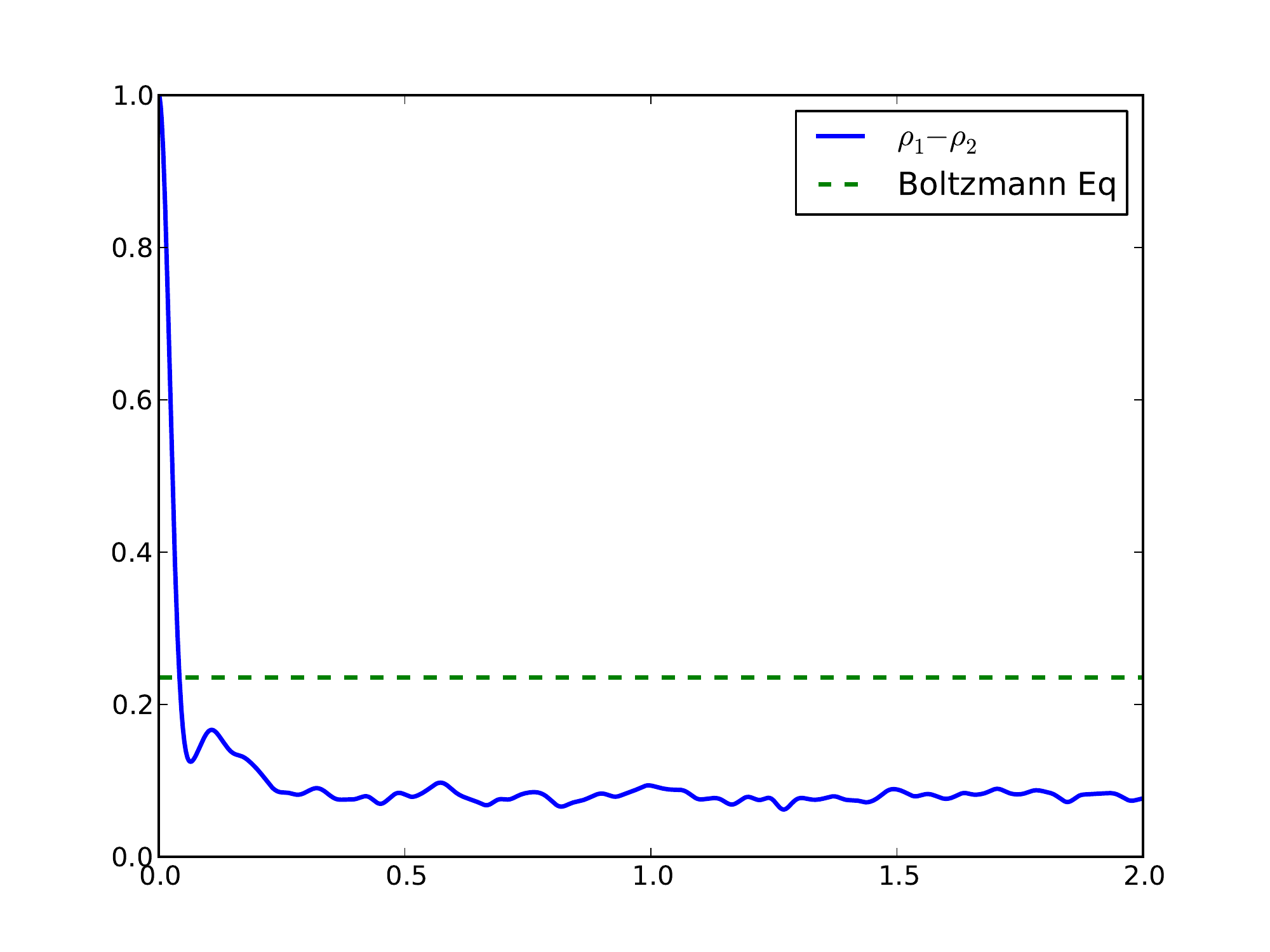}}
\end{tabular}
\caption{The reduced population dynamics, $\rho_1 - \rho_2$, in the two level system with (left) and without (right) the detailed balance correction under the Ohmic spectral density with exponential cutoff. The Boltzmann equilibrium population difference between levels 1 and 2, $\frac{\exp(-\beta \epsilon_1) - \exp(-\beta \epsilon_2)}{\exp(-\beta \epsilon_1) + \exp(-\beta \epsilon_2)}$, is the green dashed line.
}\label{2level-a}
\end{figure}

I consider a two level system to demonstrate the detailed balance correction for the modified Ehrenfest Scheme. The specifications of the two level system are $\epsilon_0=0 \text{cm}^{-1}$, $\epsilon_1=100 \text{cm}^{-1}$ and $J_{12}=J_{21}=0$ (the coherent channel is turned off). The results with and without the quantum correction are shown in Figs.~\ref{2level-a} for the phonon reservoir having a Ohmic spectral density with a exponential cutoff $\eta \omega e^{-\omega/\omega_c}$. The Ohmic spectral density has the following parameters, $\eta=10 \text{cm}^{-1}$ and $\omega_c=10 ps^{-1}$, and
\begin{equation}
V=\mathcal{V}_{12}=
\left[ \begin{array} {cc}
0 & V_{12} \\
V_{12} & 0
\end{array} \right ],
\end{equation}
where $V_{12}=1.0$. Also for this reservoir, I set temperature $T=300k$.
In Figure~\ref{2level-a}, I show the population difference of level 1 and 2, $\rho_{1}-\rho_{2}$. The initial total population is on level 1, $\rho_{11}(0)=1$ ($\vert \phi(0)\rangle =[1,\;0]^T $).
For both calculations, I use 8000 configurations. The convergence of the simulation is checked (not displayed).
Fig.~\ref{2level-a} shows that the modified Ehrenfest scheme can approach to the Boltzmann equilibrium, but the original Ehrenfest scheme can't.

\subsubsection{Three Level System Coupled to One Reservoir}\label{case2}
In this section, I consider the additional third energy level to elaborate the quantum path interference and steady state population manipulation due to the energy splitting of $\epsilon_2$ into $\epsilon_2$ and $\epsilon_3$ as shown in Figure~\ref{diagram}. The third energy level is $\epsilon_3=120\;cm^{-1}$. 

\begin{figure} \begin{tabular}{cc}
\subfloat[]{\includegraphics[width=3in]{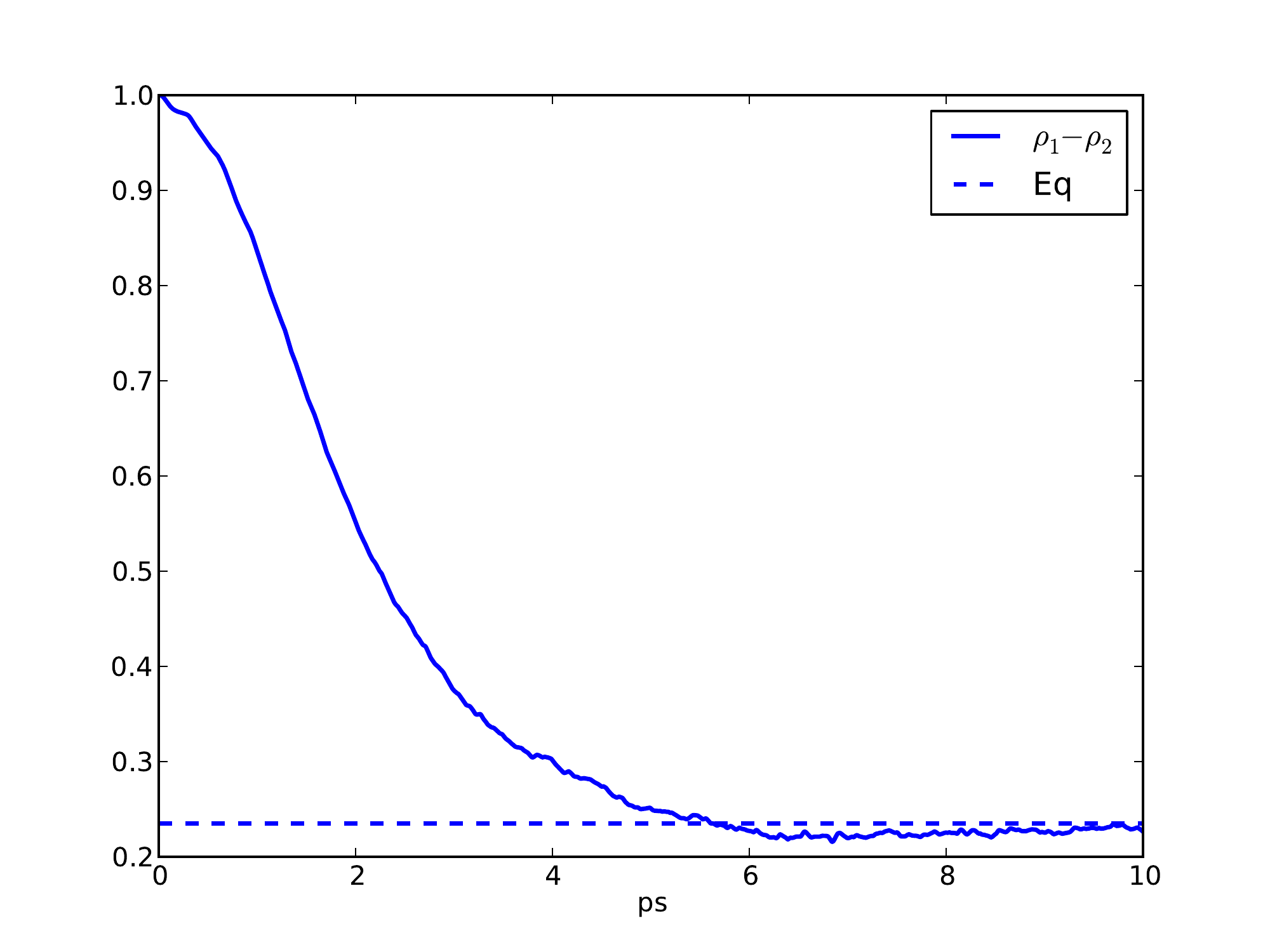}}
   & \subfloat[]{\includegraphics[width=3in]{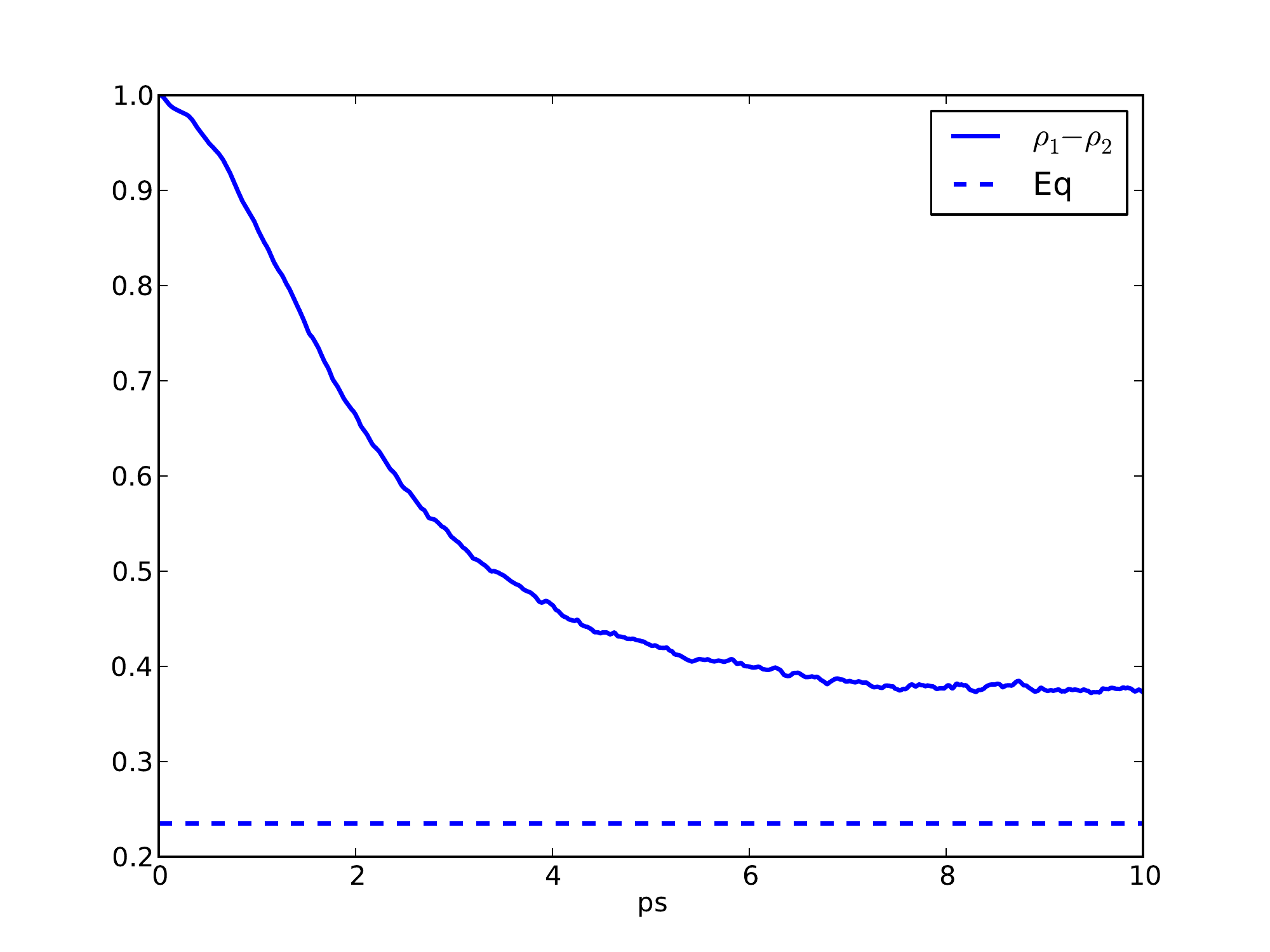}}
\end{tabular}
\centering
\subfloat[]{\includegraphics[width=3in]{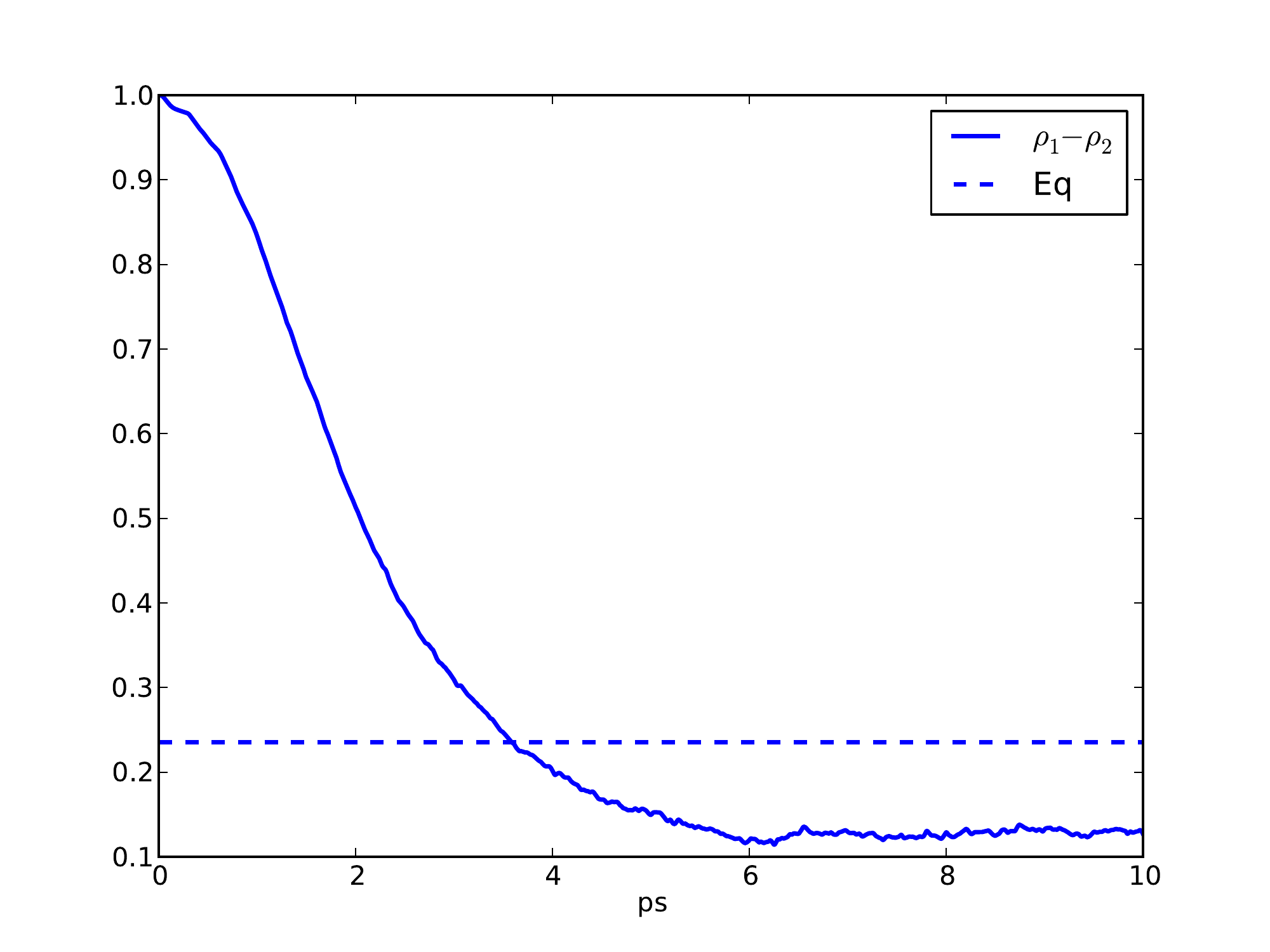}}
\caption{The reduced population dynamics, $\rho_{1}-\rho_{2}$, in the three level system under one thermal reservoir for the three cases: 1. $V_{13}=1$ and $V_{23}=1$; 2. $V_{13}=3$ and $V_{23}=1$; and 3. $V_{13}=1$ and $V_{23}=3$. The Boltzmann equilibrium population difference between levels 1 and 2, $\frac{\exp(-\beta \epsilon_1) - \exp(-\beta \epsilon_2)}{\exp(-\beta \epsilon_1) + \exp(-\beta \epsilon_2)}$, is the green dashed line}    \label{3level-a}
\end{figure}

In this setup, I have
$V=\mathcal{V}_{13} + \mathcal{V}_{23}$ where
\begin{equation}
\mathcal{V}_{13}=
\left[ \begin{array} {ccc}
0 & 0 & V_{13} \\
0 & 0 & 0 \\
V_{13} & 0 & 0
\end{array} \right ].
\end{equation}
\begin{equation}
\mathcal{V}_{23}=
\left[ \begin{array} {ccc}
0 & 0 &  0 \\
0 & 0 & V_{23}\\
0 & V_{23} & 0
\end{array} \right ].
\end{equation}
The results of the normalized difference of the steady state populations of level 1 and 2 are presented in Figure~\ref{3level-a} and  compared to the Boltzmann thermal equilibrium (green line). The normalization is defined as $\frac{\rho_1-\rho_2}{\rho_1+\rho_2}$. I consider the following three cases: 1. $V_{13}=1$ and $V_{23}=1$; 2. $V_{13}=3$ and $V_{23}=1$; and 3. $V_{13}=1$ and $V_{23}=3$. I use one reservoir in this subsubsection which is the same one used in the previous subsubsection.

With one tiny caveat, the first case among the three, $V_{13}=V_{23}=1$ is the modeled used in the literature to study the exciton transfer in the context of the Bloch-Redfield equation and the one with secular approximations \cite{Yang,berkelbach,berkelbach2}, $\it{i.e.}$, $g_k^{ij}$ are the same for the pairs $13$ and $23$. In our paper, I didn't consider the energy fluctuations, $\it{i.e.}$ $V_{ii}=0$. When $V_{13}\neq V_{23}$, I have diffident relaxations for level 1 and level 3, and level 2 and level 3. The first plot in Figure~\ref{3level-a} shows that the system can relax to the Boltzmann equilibrium as shown in the previous models\cite{berkelbach,berkelbach2}.

Figure~\ref{3level-a} shows that Case can give the Boltzmann equilibrium; but
Cases 2 and 3 can lead to the steady state population different for the Boltzmann distribution dues to the discrepancy of damping strengths $\eta$ associated with $V_{13}$ and $V_{23}$. The ratio between $\eta$ and $\epsilon_1-\epsilon_2$ is important to the quantum path interference and needs more careful study in the future.

\subsubsection{Three Level System Coupled to Two Thermal Reservoirs}\label{case3}
\begin{figure}
\includegraphics[width=5in]{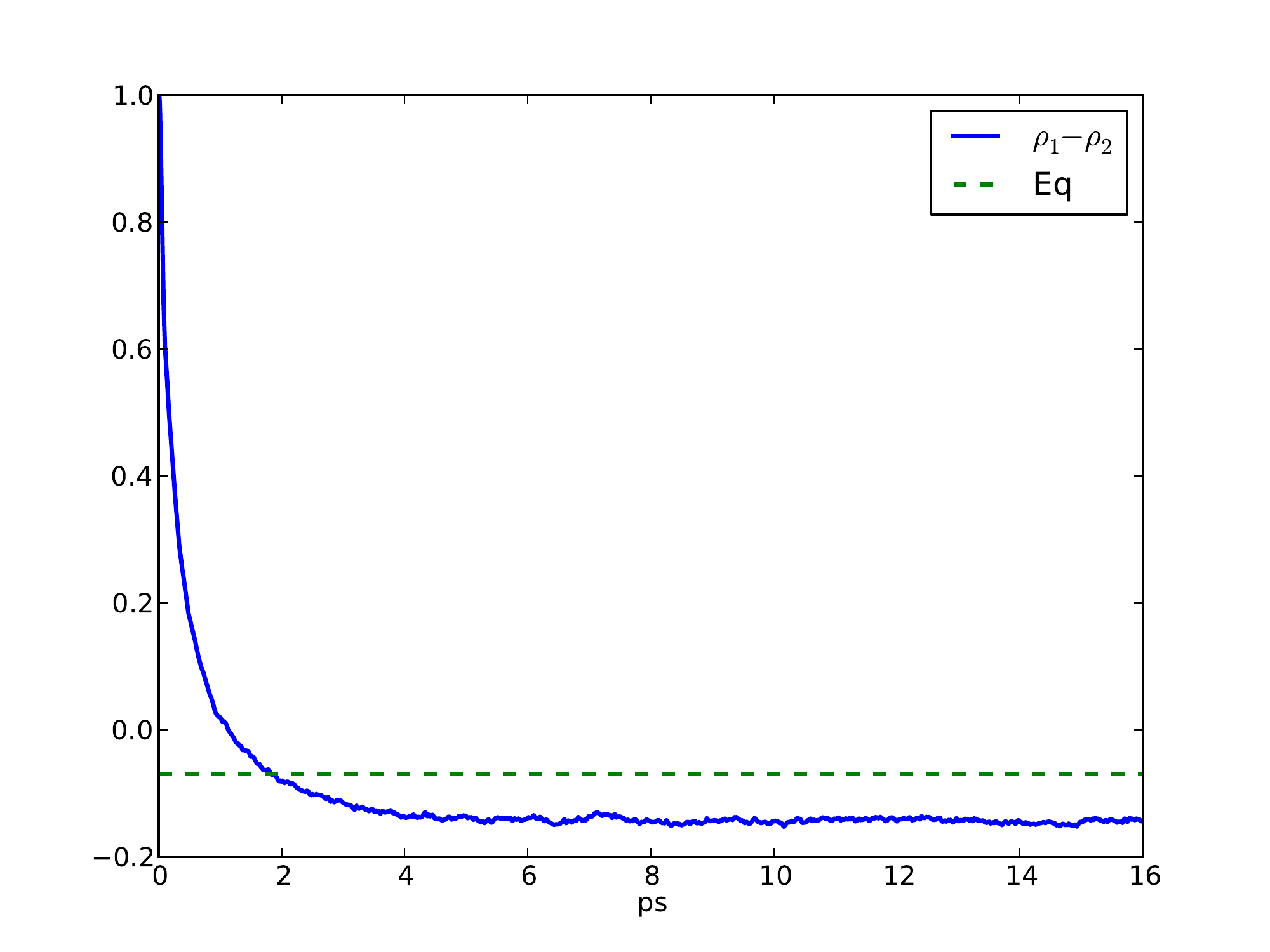}
\caption{Energy splitting at level 2 and ensuing quantum path interference can populate more on level 2 than the equilibrium proportionally. The Boltzmann equilibrium population difference under two temperature reservoirs, $\frac{\rho_1^p-\rho_2^p}{\rho_1^p+\rho_2^p}$, is the Green dashed line}\label{twotemp}
\end{figure}

In this section, I present the results for the same three level systems under two different thermal reservoirs of two different temperature. One of the thermal reservoirs can be replaced with incoherent light, particularly solar light. I use the same Ohmic spectral density as the previous subsubsection and run two separate sets of trajectories for the two thermal reservoirs. I couple the high temperature reservoir at $T=6000K$ to transition between levels 1 and 3, $\mathcal{V}_{13}\times X_{hot}$ and cold reservoir at $T=300k$  to the one between levels 2 and 3, $\mathcal{V}_{23}\times X_{cold}$.
I choose $V_{13}=1$ and $V_{23}=1$.
Figure~\ref{twotemp} shows that the energy splitting and two different temperature reservoirs can invert the steady state population away from the Boltzmann equilibrium\cite{heateng}.
Since I have two temperatures, the proportion of Boltzmann equilibrium populations will be $\rho_1^p=1$, $\rho_3^p= \rho_1 \frac{\exp(- \beta_{hot} H_3)}{\exp(- \beta_{hot} H_1)}$ and $\rho_3^p=\rho_1^p\frac{\exp(- \beta_{hot} H_3)}{\exp(- \beta_{hot} H_1)} \frac{\exp(- \beta_{cold} H_2)}{\exp(- \beta_{cold} H_3)}$. Then I normalize the three-level equilibrium population difference, $\rho_1-\rho_2=\frac{\rho_1^p-\rho_2^p}{\rho_1^p+\rho_2^p}$ for the two-level system. Figure~\ref{twotemp} shows that the quantum path interference can invert the steady state population under two different temperature reservoirs.

\section{Weak Coupling Limit and Detailed Balance} \label{hove}

Fluctuation-dissipation theory is the foundation of the non-equilibrium theory \cite{kubo}.
The Kubo-Green formulas on the linear response theory is an important bridge between the microscopic and macroscopic descriptions for the fluctuation-dissipation theory. However the theory is based on the weak coupling limit\cite{vliet}. van Kampen's objection to the linear response theory for the non-weak coupling case is an important topic for the recent study on the excitation energy transfer in the light harvesting complex\cite{engel}. But for the systems where dissipation is due to weak interactions, amenable to
the Van Hove limit, and having sufficiently short relaxation times (Markovian Limit, delta time correlation),  the Kubo-Green formulas should hold and the corresponding detailed balance determined by the two time correlation function (Fermi Golden rule as a rate at the Markovian limit) induced by the bath should be enforced.
In describing the Van Hove limit (and related Weisskopf--Wigner approximation often used in quantum optics),  the average 
effect of the interaction should be zero. Otherwise the time scale associated with reduced system is not large enough to led to microscopic fluctuations\cite{Antonio}. 

The relationship between the weak coupling limit and its detailed balance according to the linear Master equations pose a great challenge theoretically when the interaction is beyond the weak coupling limit. 
For the intermediate coupling range, the high order multi-time correlation functions (memory kernel) can contribute significantly to the path interference beyond the two-time correlation function.
For example, the multi-time correlation function of Gaussian process for phonon will have the following iterative definition\cite{Shapiro},
\begin{eqnarray}
 \langle Q(t_0)Q(t_1)\cdots Q(t_{N-1})Q(t_N) \rangle&=&\langle Q(t_0)Q(t_1)\rangle \langle Q(t_2) \cdots Q(t_N) \rangle + \\ \nonumber
 &&\langle Q(t_0)Q(t_2)\rangle \langle Q(t_1) \cdots Q(t_N) \rangle + \cdots \\ \nonumber
 &&\langle Q(t_0)Q(t_N)\rangle \langle Q(t_1) \cdots Q(t_{N-1}) \rangle.
 \end{eqnarray}
In general cases, you can not factorizing the multi-time correlation into a single product of two-time correlation function, $C(t)=\langle Q(t) Q((0) \rangle$. In order to study the quantum path interference beyond the weak coupling limit, I need to establish the non-equilibrium detailed balance according to the closed-time-path Green's function\cite{xin,Chou,Hao} and provide a complete description of the multi-time correlation function. The complex-Gaussian process constructed based on the influence functional may be a viable process\cite{xin}.

\section{Concluding Remark}\label{conclusion}

The quantum path interferences through coherent/incoherent radiative and incoherent non-radiative channels have been considered in the paper.  
I proposed an unified model to study the two channels together. 
In order to simulate the evolution of quantum subsystems with correct detailed balance, the modified Ehrenfest scheme is proposed.  I further discuss the relationship between detailed balance and weak coupling limit. The future work should consider the extension of the work with the influence functional and closed-time-path Green's function\cite{Chou} and the construction of the rigorous (complex) Gaussian process to reproduce the influence functional.

However, this method should be attractive for large scale quantum molecular dynamics simulations in the realistic open quantum systems, like solar cell, LED, organic LED, light harvesting system, $\it{etc}.$ I would like to build sophisticate realistic computational models on the top of the unified model and the current modified Ehrenfest scheme.

\section{Acknowledgment }

\bibliography{ref}

\end{document}